\documentstyle[pra,aps]{revtex}
\begin{document}
\draft
\title{
MOMENTUM AND ENERGY DISTRIBUTIONS OF NUCLEONS\\
IN FINITE NUCLEI DUE TO SHORT-RANGE CORRELATIONS}
\author{H. M\"uther}
\address{
Institut f\"ur Theoretische Physik, Universit\"at T\"ubingen,\\
Auf der Morgenstelle 14, D-72076 T\"ubingen, Germany  }
\author{A. Polls}
\address{Departament d'Estructura i Constituens de la Materia\\
Universitad de Barcelona\\
Diagonal 647, E-08028 Barcelona, Spain}
\author{W.H. Dickhoff}
\address{
Department of Physics, Washington University,\\
St. Louis, MO 63130, USA}
\date{\today}
\maketitle
\begin{abstract}
The influence of short-range correlations on the momentum and energy
distribution of
nucleons in nuclei is evaluated assuming a realistic meson-exchange potential
for the nucleon-nucleon interaction. Using the Green-function approach the
calculations are performed directly for the finite nucleus $^{16}$O avoiding
the local density approximation and its reference to studies of infinite
nuclear matter. The nucleon-nucleon correlations induced by the short-range
and tensor components of the interaction yield an enhancement of the
momentum distribution at high momenta as compared to the Hartree-Fock
description. These high-momentum components
should be observed mainly in nucleon knockout reactions like $(e,e'p)$
leaving the final nucleus in a state of high excitation energy. Our
analysis also demonstrates that non-negligible contributions to the momentum
distribution should be found in partial waves which are unoccupied in the
simple shell-model. The treatment of correlations beyond the
Brueckner-Hartree-Fock approximation also yields an improvement for the
calculated ground-state properties.
\end{abstract}

\section{Introduction}
Many properties of nuclei can be understood within the independent
particle model (IPM). In the IPM the nucleus is considered to be a system
of nucleons moving without residual interaction in a mean field or
single-particle potential.
The single-particle potential is either adjusted in
a phenomenological way (assuming e.g. a Woods-Saxon shape) or evaluated
from empirical effective interactions like the Skyrme forces within the
Hartree-Fock approximation.
Attempts to employ realistic nucleon-nucleon (NN) interactions,
which reproduce the NN scattering data, directly in such a scheme fail badly:
typically one does not
even obtain any binding energy in this approach. This result
is due to the strong
short-range and tensor components, which are typical for realistic interactions
and induce corresponding NN correlations in the nuclear wave function, which
cannot be described by the IPM or the Hartree-Fock approach.

Various tools have been developed to account for these strong
short-range correlations. These include variational calculations assuming
Jastrow correlation functions \cite{pand}, the correlated basis function
method (CBF) \cite{cbm}, the ``exponential S'' method \cite{zabo}, the
Brueckner-Hartree-Fock (BHF) approximation \cite{day} and the self-consistent
Green function approach \cite{rev}.

Considerable effort has also been made to find a nuclear property
which is experimentally accessible and reflects the effects of NN correlations
in a clear manner.
A candidate for such an observable is the momentum distribution of nucleons
in a nucleus. This momentum distribution can be written as
\begin{eqnarray}
n(k)  =& \sum_{l,j,\tau} (2j+1)\, n_{lj\tau}(k) \nonumber\\
 =& \sum_{l,j,\tau} (2j+1)\,
 \left\langle \Psi^A_0 \right\vert a^{\dag}_{klj\tau} a_{klj\tau}
\left\vert \Psi^A_0 \right\rangle .
\label{eq:nljk}
\end{eqnarray}
Here $\left\vert \Psi^A_0 \right\rangle$ represents the ground state of the
nucleus under consideration (with
A nucleons) and $a_{klj\tau }^{\dag}$ ($a_{klj\tau}$) denotes the creation
(annihilation) operator for a nucleon with orbital angular momentum $l$, total
angular momentum $j$, isospin $\tau$ and momentum $k$. The momentum
distributions for the partial waves, $n_{lj\tau}(k)$, in Eq. (\ref{eq:nljk})
can be rewritten by inserting a complete set of eigenstates
$\left\vert \Psi^{A-1}_n \right\rangle$ for the
system with $A-1$ nucleons
\begin{equation}
n_{lj\tau}(k) = \sum_n \bigl| \left\langle \Psi^{A-1}_n \right| a_{klj\tau}
\left| \Psi^A_0 \right\rangle \bigr|^2 .
\label{eq:nljke}
\end{equation}
In the IPM the sum in this equation is typically reduced to one term, if
$(l,j,\tau )$ refer to a single-particle orbit occupied in
$\left\vert \Psi^A_0 \right\rangle$.
Eq. (\ref{eq:nljke}) then yields the square of the momentum-space wave function
for this single-particle state. The contribution $n_{lj\tau}(k)$ vanishes
in the IPM if no state with quantum numbers $(l,j,\tau )$ is occupied. If
correlations are present beyond the IPM approach this simple picture is
no longer true and the determination of the momentum distribution $n(k)$
requires both in experimental as well as theoretical studies the complete
knowledge of the nucleon-hole spectral function
\begin{equation}
S_{lj\tau}(k,E) = \sum_n  \bigl| \left\langle \Psi^{A-1}_n \right| a_{klj\tau}
\left| \Psi^A_0 \right\rangle \bigr|^2 \delta (E-(E^A_0-E^{A-1}_n))
\label{eq:sljke}
\end{equation}
for all energies $E$  and all sets
of discrete quantum numbers $(l,j,\tau )$. Note that the energy variable $E$ in
this definition of the spectral function refers to the negative excitation
energy of state $n$ in the $A-1$ system with respect to the
ground-state energy of the nucleus with $A$ nucleons ($E^A_0$). The spectral
function is experimentally accessible by analyzing nucleon knockout
experiments like $(e,e'p)$. The momentum distribution $n_{lj\tau}(k)$ is
obtained by integrating the spectral function over energies $E$ from $-\infty$
to the Fermi energy $\epsilon_F = E^A_0 - E^{A-1}_0$ with $E^{A-1}_0$
denoting the energy of the ground state for $A-1$ nucleons.
One important aim of our studies is to investigate how
short-range correlations modify the spectral function at various energies
as compared to the IPM. For example, can one expect to observe effects of
short-range correlations in knockout experiments with small energy transfer?

Microscopic calculations of the spectral function and the momentum
distribution based on realistic nuclear hamiltonians have mainly been
performed for very light nuclei ($A\leq 4$) \cite{morita,tadok,ciofi,benpa}
or nuclear matter \cite{mahau,ramos,benff,ciofb,vonde,gearh,baldo,kohl}.
Results
for heavier nuclei are typically derived from investigations of nuclear
matter assuming a local density approximation \cite{string,sffb,bffs,neck}.
Recent variational calculations for $^{16}$O yield the momentum distribution
\cite{pwp} and the $p$-shell quasihole wave functions \cite{pieper}
but not the complete energy dependence of the hole spectral functions.

The present investigation determines the spectral function and the
corresponding momentum distribution directly for finite nuclei without
employing the local density approximation.
Calculations are performed for the nucleus
$^{16}$O assuming a realistic meson-exchange potential
\cite{rupr} for the NN interaction. The spectral function is derived from the
Lehmann representation of the single-particle Green function. This Green
function solves the Dyson equation with a self-energy
calculated by techniques as described in ref.\cite{boro}. A few results
concerning the spectral function for the $p_{1/2}$ partial wave have been
discussed already in a brief report \cite{brief}.

Special attention is paid to the effect of correlations on the spectral
function at different energies. We find that clear indications of the
short-range NN correlations are obtained by studying the spectral
function at very negative energies, which in nucleon knockout experiments
correspond to excitation energies of around 100 MeV and more in the
remaining nucleus. The resulting Green function is also used to study
the effects of correlations beyond the BHF approach on the binding
energy and radius of the nuclear ground state.

After this short introduction we describe the techniques used to evaluate
the spectral functions and momentum distributions in section 2. The results
of our numerical studies are presented in section 3 and the final section 4
summarizes the main conclusions of this work.

\section{Evaluation of the Spectral Functions}
The spectral function for the various partial waves, $S_{lj\tau} (k,E)$,
(see Eq. (\ref{eq:sljke})) can be obtained
from the imaginary part of the corresponding single-particle Green
function or propagator $g_{lj}(k,k;E)$. Note that here and
in the following  we have dropped the isospin quantum number $\tau$.
Ignoring the Coulomb interaction between the protons the Green functions
are identical for $N=Z$ nuclei and therefore independent of the quantum
number $\tau$. The single-particle propagator can be
obtained by solving the Dyson equation
\begin{equation}
g_{lj}(k_1,k_2;E) = g^{(0)}_{lj}(k_1,k_2;E)
+ \int dk_3\int dk_4 g^{(0)}_{lj}(k_1,k_3;E) \Delta\Sigma_{lj}(k_3,k_4;E)
g_{lj}(k_4,k_2;E) ,
\label{eq:dyson}
\end{equation}
where $g^{(0)}$ refers to a Hartree-Fock propagator and $\Delta\Sigma_{lj}$
represents contributions to the real and imaginary part of the irreducible
self-energy, which go beyond the Hartree-Fock approximation of the nucleon
self-energy used to derive $g^{(0)}$. The definition and evaluation of the
Hartree-Fock contribution as well as the calculation of $\Delta\Sigma$ are
presented in the next subsection. The methods used to solve the Dyson
equation (\ref{eq:dyson}) and to extract spectral functions as well as
momentum distributions are described in subsection 2.2.

\subsection{Nucleon Self-energy $\Sigma$}
The calculation of the self-energy is performed in terms of a $G$-matrix
which is obtained as a solution of the Bethe-Goldstone equation for
nuclear matter
\begin{eqnarray}
\left\langle k'l'SJ_SKLT \right| G & \left| k''l''SJ_SKLT \right\rangle =
\left\langle k'l'SJ_SKLT \right|
V_{NN} \left| k''l''SJ_SKLT \right\rangle   \nonumber\\ \nonumber
+ \sum_l \int k^2 dk\; & \left\langle k'l'SJ_SKLT\right| V_{NN} \left|
klSJ_SKLT\right\rangle \\
& \left\langle klSJ_SKLT \right| G \left|
k''l''SJ_SKLT \right\rangle
\frac{Q (k,K)}{\omega_{NM}- \frac{K^2}{4m} - \frac{k^2}
{2m}} .
\label{eq:betheg}
\end{eqnarray}
In this equation $k$, $k'$, and $k''$ denote the relative momentum,
$l$, $l'$, and $l''$ the orbital angular momentum for the relative motion,
$K$ and $L$ are the corresponding quantum numbers for the center of mass
motion, $S$ and $T$ denote the total spin and isospin of the interacting
pair of nucleons and by definition the angular momentum $J_S$ is obtained
from coupling the orbital angular momentum of relative motion and the spin
$S$. For the bare NN interaction $V_{NN}$ we have chosen the One-Boson-Exchange
potential $B$ defined by Machleidt (\cite{rupr}, Tab.A.2), $m$ represents
the mass of the nucleon, and the Pauli operator $Q$ is approximated by the
so-called angle-averaged approximation for nuclear matter with a
Fermi momentum $k_F = 1.4 fm^{-1}$. This roughly corresponds
to the saturation density of nuclear matter. The starting energy
$\omega_{NM}$ has been chosen to be -10 MeV. The choices for the density
of nuclear matter and the starting energy are rather arbitrary. It turns
out, however, that the calculation of the Hartree-Fock term is not very
sensitive to this choice\cite{bm2}. Furthermore, we will correct this
nuclear matter approximation by calculating the 2-particle 1-hole (2p1h)
term displayed in Fig.\ \ref{fig:diag}b
directly for the finite system, correcting the double-counting contained
in the Hartree--Fock term (see discussion below).

Using vector bracket transformation coefficients \cite{vecbr}, the $G$-matrix
elements obtained from (\ref{eq:betheg}) can be transformed from the
representation in coordinates of relative and center of mass momenta
to the coordinates of single-particle momenta in the laboratory frame in which
the 2-particle state would be described by quantum numbers such as
\begin{equation}
\left| k_1 l_1 j_1 k_2 l_2 j_2 J T \right\rangle ,  \label{eq:bra2p}
\end{equation}
where $k_i$, $l_i$ and $j_i$ refer to momentum and angular momenta of
particle $i$ whereas $J$ and $T$ define the total angular momentum and
isospin of the two-particle state. It should be noted that Eq. (\ref{eq:bra2p})
represents an antisymmetrized 2-particle state.
Performing an integration over one of
the $k_i$, one obtains a 2-particle state in a mixed representation of one
particle in a bound harmonic oscillator while the other is in a
plane wave state
\begin{equation}
\left| n_1 l_1 j_1 k_2 l_2 j_2 J T \right\rangle = \int_0^\infty dk_1 \, k_1^2
R_{n_1, l_1}(\alpha k_1) \, \left| k_1 l_1 j_1 k_2 l_2 j_2 J T\right\rangle
..
\label{eq:branp}
\end{equation}
Here $R_{n_1, l_1}$ stands for the radial oscillator function and the
oscillator length $\alpha = 1.72 $ fm$^{-1}$ has been selected. This
choice for the oscillator length corresponds to an oscillator energy of
$\hbar \omega_{osc}$ = 14 MeV. Therefore the oscillator functions
are quite appropriate to describe the wave functions of the bound
single-particle states in $^{16}$O. Using the nomenclature defined in
Eqs. (\ref{eq:betheg}) - (\ref{eq:branp}) our Hartree-Fock approximation
for the self-energy is easily obtained in the momentum representation
\begin{equation}
\Sigma^{HF}_{l_1j_1} (k_1,k'_1) =
\frac{1}{2(2j_1+1)} \sum_{n_2 l_2 j_2 J T} (2J+1) (2T+1)
\left\langle k_1 l_1 j_1 n_2 l_2 j_2 J T \right| G \left|
k'_1 l_1 j_1 n_2 l_2 j_2 J T\right\rangle .
\label{eq:selhf}
\end{equation}
The summation over the oscillator quantum numbers is restricted to the
states occupied in the IPM of $^{16}$O. This
Hartree-Fock part of the self-energy is real and does not depend on the
energy.

The terms of lowest order in $G$ which give rise to an imaginary part
in the self-energy are represented by the diagrams displayed in Figs.\
\ref{fig:diag}b) and \ref{fig:diag}c),
refering to intermediate 2p1h and 2-hole 1-particle (2h1p)
states respectively. The 2p1h contribution to the imaginary part is given
by
\begin{eqnarray}
{W}^{2p1h}_{l_1j_1} (k_1,k'_1; E)
=& \frac{-1}{2(2j_1+1)}  \sum_{n_2 l_2 j_2} \sum_{l L}
\sum_{J J_S S T} \int k^2 dk \int K^2 dK  (2J+1) (2T+1) \nonumber \\
& \times \left\langle k_1 l_1 j_1 n_2 l_2 j_2 J T \right| G \left|
k l S J_S K L T \right\rangle
  \left\langle k l S J_S K L T \right| G \left| k'_1 l_1 j_1 n_2 l_2 j_2 J T
  \right\rangle \nonumber \\
& \times \pi \delta\left(E +\epsilon_{n_2 l_2 j_2}
-\frac{K^2}{4m}-\frac{k^2}{m}\right), \label{eq:w2p1h}
\end{eqnarray}
where the ``experimental'' single-particle energies $\epsilon_{n_2 l_2 j_2}$
are used for the hole states, while the energies of the particle states are
given in terms of the kinetic energy only.
The expression in Eq. (\ref{eq:w2p1h}) still ignores the requirement that the
intermediate particle states must  be orthogonal to the hole states, which
are occupied for the nucleus under consideration. The techniques to incorporate
the orthogonalization of the intermediate plane wave states to the occupied
hole states are discussed in detail by Borromeo et al.\cite{boro}.
The 2h1p contribution to the imaginary part ${W}^{2h1p}_{l_1j_1}(k_1,k'_1; E)$
can be calculated in a similar way (see also \cite{boro}).

Our choice to assume pure kinetic energies for the particle states in
calculating the imaginary parts of $W^{2p1h}$ (Eq. (\ref{eq:w2p1h})) and
$W^{2h1p}$ may not be very realistic for the excitation modes at low energy.
Indeed a sizeable imaginary part in $W^{2h1p}$ is obtained only for
energies $E$ below -40 MeV. As we are mainly interested, however, in the
effects of short-range correlations, which lead to excitations of particle
states with high momentum, the choice seems to be appropriate.
A different approach would be required to treat the coupling
to the very low-lying two-particle-one-hole and two-hole-one-particle
states in an adequate way. Attempts at such a treatment can be found in
Refs.\ \cite{brand,rijsd,skou1,skou2}.

The 2p1h contribution to the real part of the self-energy can be calculated
from the imaginary part $W^{2p1h}$ using a dispersion relation
\cite{mbbd}
\begin{equation}
V^{2p1h}_{l_1j_1}(k_1,k_1';E)=\frac{P}{\pi} \int_{-\infty}^{\infty} \frac
{W^{2p1h}_{l_1j_1}(k_1,k_1';E')}{E'-E} dE', \label{eq:disper1}
\end{equation}
where $P$ means a principal value integral. From the $\delta$-function in
Eq. (\ref{eq:w2p1h}) one can see
that $W^{2p1h}$ is different from zero only for positive values of
$E'$. Since the diagonal matrix elements are negative, the
dispersion relation (\ref{eq:disper1}) implies that the diagonal elements of
$V^{2p1h}$ will be attractive for negative energies $E$.
They will decrease and change sign only for large
positive values for the energy of the interacting nucleon.

Since the Hartree--Fock contribution $\Sigma^{HF}$
has been calculated in terms of a
nuclear matter $G$-matrix, it already contains 2p1h terms of the kind displayed
in Fig.\ \ref{fig:diag}b).
Therefore one would run into problems of doublecounting if one
simply adds the real part $V^{2p1h}$ to the Hartree-Fock self-energy. Notice
that $\Sigma^{HF}$ does not contain any imaginary part because it is calculated
with a nuclear matter $G$-matrix at a starting energy for which $G$ is real.
In order to avoid such an overcounting of the particle-particle
ladder terms, we subtract from the real part of the self-energy a
correction term, which just contains this contribution calculated in nuclear
matter. This correction $V_c$
is given by
\begin{eqnarray}
V_c(k_1, k'_1) = \frac{1}{4(2j_1+1)} &\sum_{n_2 l_2 j_2 } \sum_{l L }
\sum_{J J_S S T} \int k^2 dk \int K^2 dK (2J+1) (2T+1) \nonumber\\
& \times \left\langle k_1 l_1 j_1 n_2 l_2 j_2 J T \right| G
\left| k l S J K L T \right\rangle
  \left\langle k l S J K L T \right| G
  \left| k'_1 l_1 j_1 n_2 l_2 j_2 J T \right\rangle \nonumber\\
& \times \frac{Q(k,K)}{\omega_{NM}
-\frac{K^2}{4m } -\frac{k^2}{m}},\label{eq:corr}
\end{eqnarray}
with the same starting energy $\omega_{NM}$ and the Pauli operator $Q$ as
used in the Bethe-Goldstone equation (\ref{eq:betheg}).

A dispersion relation similar to Eq. (\ref{eq:disper1})
holds for $V^{2h1p}$ and $W^{2h1p}$
\begin{equation}
V^{2h1p}_{l_1j_1}(k_1,k_1';E)=-\frac{P}{\pi} \int_{-\infty}^{\infty} \frac
{W^{2h1p}_{l_1j_1}(k_1,k_1';E')}{E'-E} dE', \label{eq:disper2}
\end{equation}
Since $W^{2h1p}$ is positive (at least its diagonal matrix elements) and
different from zero for negative energies $E'$ only, it is evident
from Eq. (\ref{eq:disper2}) that $V^{2p1h}$ is repulsive for positive
energies and decreases
with increasing energy. Only for large negative energies it becomes
attractive.

Summing up the various contributions we obtain for the self-energy the
following expressions
\begin{eqnarray}
\Sigma & = \Sigma^{HF} + \Delta\Sigma \nonumber\\
&  = \Sigma^{HF} + \left( V^{2p1h} - V_c + V^{2h1p}\right)
+ i\left( W^{2p1h} + W^{2h1p} \right) . \label{eq:defsel}
\end{eqnarray}

\subsection{Solution of the Dyson equation}

After we have determined the various contributions to the nucleon self-energy,
we now want to solve the Dyson equation (\ref{eq:dyson}) for the
single-particle propagator.
In order to discretize the integrals in this equation we consider
a complete basis within a spherical box of a radius $R_{\rm box}$. This box
radius should be larger than the radius of the nucleus considered.
The calculated observables are independent of the choice of $R_{\rm box}$,
if it is chosen to be around 15 fm or larger. A complete and
orthonormal set of regular basis functions within this box is given by
\begin{equation}
\Phi_{iljm} ({\bf r}) = \left\langle {\bf r} \vert k_i l j m
\right\rangle = N_{il} j_l(k_ir)
{\cal Y}_{ljm} (\theta\phi ) \label{eq:boxbas}
\end{equation}
In this equation ${\cal Y}_{ljm}$ represent the spherical harmonics
including the spin degrees of freedom and $j_l$ denote the spherical
Bessel functions for the discrete momenta $k_i$ which fulfill
\begin{equation}
j_l (k_i R_{\rm box}) = 0 .
\label{eq:bound}
\end{equation}
Using the normalization constants
\begin{equation}
N_{il} =\cases{\frac{\sqrt{2}}{\sqrt{R_{\rm box}^3} j_{l-1}(k_i
R_{\rm box})}, & for $l > 0$ \cr
\frac{i \pi\sqrt{2}}{\sqrt{R_{\rm box}^3}}, & for $l=0$,\cr}
\label{eq:nbox}
\end{equation}
the basis functions defined in Eq. (\ref{eq:boxbas}) are orthogonal and
normalized within the box
\begin{equation}
\int_0^{R_{\rm box}} d^3 r\, \left\langle k_{i'} l' j' m' \vert {\bf r}
\right\rangle \left\langle {\bf r}
\vert k_i l j m \right\rangle
= \delta_{ii'} \delta_{ll'} \delta_{jj'} \delta_{mm'} .
\label{eq:boxi}
\end{equation}
Note that the basis functions defined for discrete values of the momentum
$k_i$ within the box differ from the plane wave states defined in the
continuum with the corresponding momentum just by the normalization constant,
which is $\sqrt{2/\pi}$ for the latter. This enables us to determine the
matrix elements of the nucleon self-energy in the basis of Eq.
(\ref{eq:boxbas})
from the results presented in the preceeding section.

As a first step we determine the Hartree-Fock approximation for the
single-particle Green function in the ``box-basis.'' For that purpose the
Hartree-Fock Hamiltonian is diagonalized
\begin{equation}
\sum_{n=1}^{N_{\rm max}} \left\langle k_i \right| \frac{k_i^2}{2m}\delta_{in} +
\Sigma^{HF}_{lj} \left| k_n \right\rangle \left\langle k_n \vert \alpha
\right\rangle_{lj} = \epsilon^{HF}_{\alpha
lj} \left\langle k_i \vert \alpha\right\rangle_{lj}. \label{eq:hfequ}
\end{equation}
Here and in the following the set of basis states in the box has been
truncated by assuming an appropriate $N_{\rm max}$. In the basis of
Hartree-Fock states $\left| \alpha \right\rangle$, the Hartree-Fock propagator
is diagonal and given by
\begin{equation}
g_{lj}^{(0)} (\alpha; E) = \frac{1}{E-\epsilon^{HF}_{\alpha lj} \pm
i\eta} , \label{eq:green0}
\end{equation}
where the sign in front of the infinitesimal imaginary quantity $i\eta$ is
positive (negative) if $\epsilon^{HF}_{\alpha lj}$ is above (below) the
Fermi energy. With these ingredients one can solve the Dyson equation
(\ref{eq:dyson}). One possibility is to determine first the
so-called reducible self-energy, originating from an iteration of
$\Delta\Sigma$, by solving
\begin{equation}
\left\langle \alpha \right| \Sigma^{red}_{lj}(E) \left| \beta \right\rangle =
\left\langle \alpha \right| \Delta\Sigma_{lj}(E) \left| \beta \right\rangle
+ \sum_\gamma
\left\langle \alpha \right| \Delta\Sigma_{lj}(E) \left| \gamma \right\rangle
g_{lj}^{(0)} (\gamma ; E) \left\langle \gamma \right| \Sigma^{red}_{lj}(E)
\left| \beta \right\rangle
\end{equation}
and obtain the propagator from
\begin{equation}
g_{lj} (\alpha ,\beta ;E ) = g_{lj}^{(0)} (\alpha ;E )
\left\langle \alpha \right| \Sigma^{red}_{lj}(E) \left| \beta \right\rangle
g_{lj}^{(0)} (\beta ;E ) .
\end{equation}
Using this representation of the Green function one can calculate the
spectral function in the ``box basis'' from
\begin{equation}
\tilde S_{lj}^c (k_i , E) = \frac{1}{\pi} \mbox{Im} \left(
\sum_{\alpha, \beta} \left\langle k_i \vert \alpha \right\rangle_{lj} g_{lj}
(\alpha ,\beta ;E )
\left\langle \beta \vert k_i \right\rangle_{lj}\right) . \label{eq:skob}
\end{equation}
For energies $E$ below the lowest single-particle energy of a
given Hartree-Fock state (with $lj$)
this spectral function is different from zero
only due to the imaginary part in $\Sigma^{red}$.
This contribution involves the coupling to the continuum of 2h1p states
and is therefore non-vanishing only for energies at which the corresponding
irreducible self-energy $\Delta\Sigma$ has a non-zero imaginary part. Besides
this continuum contribution, the hole spectral function
also receives contributions from the quasihole states \cite{rev}. The
energies and wavefunctions of these quasihole states can be determined
by diagonalizing the Hartree-Fock Hamiltonian plus $\Delta\Sigma$ in the
``box basis''
\begin{equation}
\sum_{n=1}^{N_{\rm max}} \left\langle k_i \right| \frac{k_i^2}{2m}\delta_{in} +
\Sigma^{HF}_{lj} + \Delta\Sigma_{lj} (E=\epsilon^{qh}_{\Upsilon lj})
\left| k_n \right\rangle \left\langle k_n \vert \Upsilon \right\rangle_{lj} =
\epsilon^{qh}_{\Upsilon lj} \left\langle k_i \vert \Upsilon \right\rangle_{lj}
.. \label{eq:qhequ}
\end{equation}
Since in the present work $\Delta\Sigma$ only contains a sizeable imaginary
part for energies $E$ below $\epsilon^{qh}_{\Upsilon}$, the
energies of the quasihole states come out real and the continuum contribution
to the spectral function is separated in energy from the quasihole
contribution. The quasihole contribution to the hole spectral function is
given by
\begin{equation}
\tilde S^{qh}_{\Upsilon lj} (k_i, E) = Z_{\Upsilon lj} \left|
{\left\langle k_i \vert \Upsilon \right\rangle_{lj} }
\right| ^2 \, \delta (E - \epsilon^{qh}_{
\Upsilon lj}), \label{eq:skoqh}
\end{equation}
with the spectroscopic factor for the quasihole state given by \cite{rev}
\begin{equation}
Z_{\Upsilon lj} =
\bigg( {1-{\partial \left\langle \Upsilon \right| \Delta\Sigma_{lj}(E)
\left| \Upsilon \right\rangle \over
\partial E} \bigg|_{\epsilon^{qh}_{\Upsilon lj}}} \bigg)^{-1} .
\label{eq:qhs}
\end{equation}
Finally, the continuum contibution of Eq. (\ref{eq:skob}) and the quasihole
parts of Eq. (\ref{eq:skoqh}), which are obtained in the basis of box states,
can be added and renormalized to obtain the spectral function in the
continuum representation at the momenta defined by Eq. (\ref{eq:bound})
\begin{equation}
S_{lj} (k_i,E) = \frac{2}{\pi} \frac{1}{N_{il}^2} \bigl( \tilde S^c_{lj}
(k_i,E) + \sum_{\Upsilon} \tilde S^{qh}_{\Upsilon lj} (k_i,E) \bigr).
\label{eq:renor}
\end{equation}

\subsection{Ground-state Properties}

The single-particle propagator calculated by the techniques
described above, may also be used to evaluate expectation
values of single-particle operators, like the mean square radius, and the
energy of the ground state. For that purpose one
also needs the non-diagonal part of the density matrix, which is given in
the ``box basis,'' defined in the preceeding subsection, by
\begin{equation}
\tilde n_{lj} (k_i, k_n) = \int_{-\infty}^{\epsilon_F} dE\,
\frac{1}{\pi} \mbox{Im} \left(
\sum_{\alpha, \beta} \left\langle k_i \vert \alpha \right\rangle_{lj}
g_{lj} (\alpha ,\beta ;E )
\left\langle \beta \vert k_n \right\rangle_{lj}\right) , \label{eq:ndia}
\end{equation}
and contains, as before in the case of the spectral function, a continuous
contribution and a part originating from the quasihole states
\begin{equation}
\tilde n_{lj}^{qh} (k_i, k_n) = \sum_{\Upsilon} Z_{\Upsilon lj}
\left\langle k_i \vert \Upsilon \right\rangle_{lj}
\left\langle \Upsilon \vert k_n \right\rangle_{lj} . \label{eq:nqh}
\end{equation}
The sum in this equation is restricted to quasihole states with energies
below the Fermi energy $\epsilon_F$. With this density matrix the expectation
value for the square of the radius can be calculated according to
\begin{equation}
\left\langle \Psi^A_0 \right| r^2 \left| \Psi^A_0 \right\rangle =
\sum_{l,j} 2(2j+1)\sum_{i,n=1}^{N_{\rm max}}
\left\langle k_i \right| r^2 \left| k_n \right\rangle_l n_{lj} (k_i, k_n ) ,
\label{eq:rsqu}
\end{equation}
with a factor of 2 accounting for isospin degeneracy.
The matrix elements for $r^2$ are given by
\begin{equation}
\left\langle k_i \right| r^2 \left| k_n \right\rangle_l =
N_{il} N_{nl} \int_0^{R_{\rm box}} dr\, r^4 \,
j_l(k_ir)j_l(k_nr)\; .
\end{equation}
In the same way one can also calculate the expectation value for the
particle number.
The total energy of the ground state is obtained from the ``Koltun
sum rule''
\begin{equation}
E_0^A = \sum_{l,j} 2(2j+1)\sum_{i=1}^{N_{\rm max}}
\int_{-\infty}^{\epsilon_F} dE\,
\frac{1}{2} \left( \frac {k_i^2}{2m} + E \right) \bigl( \tilde S^c_{lj}
(k_i,E) + \sum_{\Upsilon} \tilde S^{qh}_{\Upsilon lj} (k_i,E) \bigr)\; .
\label{eq:koltun}
\end{equation}
As in Eq. (\ref{eq:renor}), the sum over quasihole states $\Upsilon$
is restricted to those below $\epsilon_F$.

\section{Results and Discussion}

In our discussion of the hole spectral function in the preceeding section
we have distinguished the contributions originating from the quasihole
states and the continuum of 2h1p configurations (see Eq. (\ref{eq:renor})).
This separation into the two parts is displayed in Fig.\ \ref{fig:specpw}
for the
energy integrated spectral function (including all energies below the Fermi
energy $\epsilon_F$) for different partial waves $lj$.
This figure displays quite
clearly that the momentum distribution at small momenta is dominated by the
quasihole contribution (for those partial waves for which it exists)
whereas the high-momentum components are given by the continuum part (see also
Ref.\ \cite{brief}).

This implies that a nucleon knockout reaction with small energy transfer,
leaving the remaining nucleus (e.g. $^{15}$N in the present case as
all results presented here refer to $^{16}$O) in its ground state or in
the lowest state with angular momentum and parity defined by the partial
wave quantum numbers $j$ and $l$, should display a spectral distribution
as presented by the quasihole part. The high-momentum components of the
spectral function (or momentum distribution) should only be observed
in experiments which also include knockout processes into states represented
by the 2h1p continuum.
We recall that the present approach has been designed to account for the
effects of short-range correlations.
Effects due to configuration mixing of the
hole state with 2h1p configuration at low energies must be treated in terms
of shell-model configuration mixing or by techniques as discussed in
Refs.\ \cite{brand,rijsd,skou1,skou2}.

In order to characterize the energy dependence of the spectral functions
one may define a mean value for the energy of the 2h1p continuum for
each momentum and each partial wave by
\begin{equation}
{\cal E}_{lj} (k) = \frac{\int dE\,E \,S_{lj}^{c} (k,E)}{\int dE
\,S_{lj}^{c} (k,E)}\;.  \label{eq:emean}
\end{equation}
Typical values for this mean value range from -80 to -150 MeV for the
momenta $k$ considered in this analysis ($k\leq 3.3$ fm$^{-1}$).
One also finds that this mean value is quite independent of the partial wave
considered (see left part of Fig.\ \ref{fig:emean}).
Therefore it is useful to define
a mean value of the energy by averaging over all
partial waves
\begin{equation}
{\cal E} (k) = \frac{\sum_{l,j}(2j+1)\,\int dE\,E \,S_{lj}^{c} (k,E)}{
\sum_{l,j}(2j+1)\ \int dE\,S_{lj}^{c} (k,E)}\;.  \label{eq:emea2}
\end{equation}
The resulting energy spectrum ${\cal E}(k)$ is shown in the left part
of Fig.\ \ref{fig:emean} and compared to a simple parametrisation of
this curve in terms of $-k^2 /(2m^*) - C$ with $m^*$ = 2400 MeV and $C$ = 80
MeV. This parametrisation demonstrates that the momentum dependence of
this mean value is weak as compared e.g. to the kinetic energy.
One may also compare
the mean value ${\cal E}(k)$ determined by
Eq. (\ref{eq:emea2}) in $^{16}$O with the corresponding quantity
obtained for nuclear matter using the Reid potential \cite{vonde,polls}.
The mean value calculated for nuclear matter shows a stronger momentum
dependence and therefore, at high momenta, yields energies considerably
below those displayed in Fig.\ \ref{fig:emean}. This implies that the
nuclear matter calculation exhibits a larger probability to excite
2p1h configurations at higher energies as compared to the present
approach. We will come back to a discussion of possible differences
between the present calculation and studies in nuclear matter when we
analyze the results for the momentum distribution below.

In order to show the importance of the continuum part of the spectral
functions as compared to the quasihole contribution and to visualize
the effects of correlations, we have included in Tab.\ref{tab:noc}
the particle numbers for each partial wave including the degeneracy of the
states
\begin{equation}
{\hat n_{lj}} = 2 (2j+1) \int_{-\infty}^{\epsilon_F}
dE\, \int_0^{\infty} dk\,k^2
S_{lj}(k,E) \;, \label{eq:nocn}
\end{equation}
also separating the contributions originating from the quasihole states
and those due to the continuum.

In the present approach the quasihole states, which in a Hartree-Fock
approximation would be occupied with a probability of 1.0, are occupied
with a probability of 0.78, 0.91 and 0.90 in the case of $s_{1/2}$,
$p_{3/2}$ and $p_{1/2}$, respectively. This means that only 14.025 out
of the 16 nucleons of $^{16}$O occupy the quasihole states. Another 1.13
``nucleons'' are found in the 2h1p continuum with partial wave quantum
numbers of the $s$ and $p$ shell, while an additional 0.687 ``nucleons''
are obtained from the continuum with orbital quantum numbers of the $d$
and $f$ shells. The distinction between quasihole and continuum
contributions is somewhat artificial for the $s_{1/2}$ orbital since
the coupling to low-lying 2h1p states leads to a strong fragmentation of the
strength \cite{domit}, which is also observed experimentally \cite{moug}.
A recent $(e,e'p)$ experiment on $^{16}$O \cite{leus} has provided detailed
information on the spectroscopic factors at low-energy transfer.
The analysis of the experiment indicates that {\it e.g.} the $p_{1/2}$
quasihole state carries only 63\% of the strength.
This result should be compared to the 90\% obtained here.
This discrepancy is partly due to the emphasis in the present work on
the accurate treatment of short-range correlations. Long-range (low-energy)
correlations, not considered in this work, typically yield another 10\%
reduction of the quasihole strength \cite{brand,rijsd,skou1,skou2,domit}.
It has also been observed that a correct treatment of the center of mass
motion may be responsible for another 10\% reduction in the quasihole
strength \cite{pieper}.

The sum of the particle numbers listed in Tab.\ref{tab:noc} is
slightly smaller (15.841) than the particle number corresponding to
$^{16}$O.
There are several possible sources for this discrepancy:
First of all our analysis only
accounts for momenta $k$ below 3.3 fm$^{-1}$ and we did not consider
partial waves with $l >3$. The restriction in $k$ is determined by the
choice of $N_{\rm max}$ in truncating the ``box basis''
(see e.g.\ Eq. (\ref{eq:hfequ})).
Inspecting the decrease of the occupation
numbers listed in Tab.\ref{tab:noc} with increasing $l$ one can
expect that the ``missing'' nucleons may be found in partial waves
with $l>3$. Furthermore, however, one must keep in mind that the
present approach to the single-particle Green function is not
number-conserving, as the Green functions used to evaluate the self-energy
are not determined in a self-consistent way \cite{rev}.
It should be pointed out that the depletion of the occupation
probabilities of the hole states, indicated in Tab.\ref{tab:noc},
is particularly large for the $s_{1/2}$ orbit.
This feature can be ascribed to the closeness of the $s_{1/2}$
Hartree-Fock energy to the 2h1p continuum which yields more leakage of
strength to the continuum than for the $p_{1/2}$ and $p_{3/2}$ quasihole
states.

Inspecting the contributions to $\hat n_{lj}^c$ originating from the various
energy regions in Tab.\ref{tab:noc}, one can see that the major
contributions are obtained from energies around -100 MeV. Only small
contributions come from energies below -150 MeV. The same feature is
also obtained if one analyzes the momentum integrated spectral function of
the continuum
\begin{equation}
N^c (E) = \sum_{l,j} 2 (2j+1) \int_0^\infty dk\, k^2 \, S^c_{lj} (k,E)
\; , \label{eq:nce}
\end{equation}
shown in Fig.\ref{fig:nce}.
As a function of the energy of the 2h1p states, this density of states
rises very rapidly just below our threshold
for 2h1p configurations at $\approx -$40 MeV, shows a maximum at -60 MeV,
a second local maximum around -85 MeV, reflecting possibly some shell
structure, and smoothly vanishes at lower energies. This density of states
corresponds to a prediction of the total spectral strength to be observed
in knockout reactions as a function of the energy transfer.

The contribution of the 2h1p continuum to the momentum distribution
is presented in Fig.\ \ref{fig:npw}, exhibiting the contributions
from partial waves with various $l$. The momentum distributions displayed
in this figure contain the degeneracy factors $2(2j+1)$ and are normalized
in such a way that $\int dk n_l(k)$ yields the total number of particles
with orbital angular momentum $l$ in the 2p1h continuum. This figure
also shows that the largest contributions are obtained for $l=0$, although
the degeneracy factor is small in this case. One can see, however,
that the decrease of the contributions with increasing $l$ is slow, supporting
the above argument that the missing particle number exhibited in
Tab.\ref{tab:noc} should be obtained from partial waves with $l>3$. In
addition,
the centroid of the momentum distribution is shifted to
higher momenta with increasing $l$. At momenta $k \approx 3$ fm$^{-1}$ the
largest contribution is obtained from $l=3$.

The total momentum distribution, including the contribution from the
quasihole states is shown in Fig.\ \ref{fig:nto1}. This distribution is
presented for various energy cut-offs. The quasihole part reflects
the cross section for knockout reactions with small energy transfer, i.e.
leading to the ground state of the final nucleus and excited states
up to $\approx 20$ MeV. The curve denoted by $E>$ -100 MeV should reflect
the momentum distribution including all states of the final
nucleus up to around 80 MeV, etc. As has been discussed already in
connection with the spectral functions of Fig.\ \ref{fig:specpw} (see
also Ref.\ \cite{brief}), the high-momentum components of the momentum
distribution due to short-range correlations are expected to be
observable mainly in
knockout experiments with an energy transfer of the order of 100 MeV.

The total momentum distributions resulting from the quasihole states and
the 2h1p continuum are displayed again in Fig.\ \ref{fig:nto2} and
compared to predictions from studies in nuclear matter \cite{vonde,polls}.
In order to enable the comparison with the nuclear matter results, the
momentum distributions resulting from the present studies have been divided
by the particle number and are normalized in this figure such that $\int
d^3k\, n(k)$ yields 1. This comparison demonstrates that the enhancement
of the momentum distribution predicted by the present study for high
momenta is well below the corresponding prediction derived from nuclear
matter.

At first sight this discrepancy seems to be in contradiction to the success of
the Local Density Approximation found in Ref.\ \cite{bffs}. Before we reach
this conclusion, however, one must consider the following points: (i)
Unfortunately, we cannot compare the momentum distribution obtained in
our study of $^{16}$O using the OBE potential $B$ of \cite{rupr} with a
momentum distribution derived for the same interaction in nuclear matter.
The curve displayed in Fig.\ \ref{fig:nto2} has been evaluated for the
Reid soft-core potential. The modern OBE potentials are considered to be
``softer'' than the older Reid potential. Therefore part of the discrepancy
might be explained by the different interaction. It would be very useful
to pursue whether various realistic interactions, indeed, predict differences
in the momentum distribution, which might be observed in experiment. (ii)
The momentum distribution of nuclear matter has been evaluated for the
empirical saturation density. In order to compare with a momentum
distribution of a light nucleus, like $^{16}$O, the momentum distribution
of nuclear matter at around half the saturation density would be more
appropriate. The momentum distribution of nuclear matter tends to
be smaller at high-momentum transfers for smaller densities \cite{wiri}.
(iii) In our present study of finite nuclei we only consider contributions to
the self-energy of the nucleons up to second order in the $G$-matrix (see
Fig.\ \ref{fig:diag}), whereas the study in nuclear matter accounts for a
self-consistent treatment of all ladder diagrams.
It is possible that a perturbative approach
underestimates the high momentum components in the distribution, since
the $G$-matrix is soft as compared to the bare potential. (iv) Our present
approach underestimates the effect of low-energy excitations (see discussion
of the single-particle spectrum in calculating the self-energy following
Eq. (\ref{eq:w2p1h})). For a finite nucleus it is quite possible that an
enhancement of these correlations due to low-energy excitations will
provide an enhancement of the momentum distribution around $k$ = 3 fm$^{-1}$.
(v) Finally, we would like to recall that partial waves with $l>3$, which
were ignored in the present study may provide a non-negligible contribution
to the momentum distribution at high momenta (see also Fig.\ \ref{fig:npw}).

Finally, we would like to discuss the effects of correlations which are
taken into account in the present investigation beyond the BHF approximation,
on the ground-state properties of $^{16}$O. For that purpose Tab.\ref{tab:kol}
lists the ingredients for calculating the total energy of the ground state
according to Eq. (\ref{eq:koltun}). Furthermore we present
results obtained for the radius of the nucleon distribution (see
Eq. (\ref{eq:rsqu})).

As a first approximation we consider the Hartree-Fock (HF) approximation,
which means that the self-energy of the nucleons is approximated by
Eq. (\ref{eq:selhf}). This implies that the occupation probabilities are
equal to 1 for the three hole states $s_{1/2}$, $p_{3/2}$, $p_{1/2}$ and
0 otherwise.
The resulting binding energy per nucleon (-1.93 MeV) is quite small.
We believe that this small binding energy is due to the use of the nuclear
matter $G$-matrix calculated at the saturation density, which overestimates
the Pauli effects as compared to a BHF calculation directly for $^{16}$O.

The treatment of the Pauli operator is improved by
adding the 2p1h part (Eq. \ref{eq:disper1}) minus the
correction term of Eq. (\ref{eq:corr}) to the self-energy, an approximation
which we will call Brueckner-Hartree-Fock (BHF) in the following. Note
that the occupation probabilities of the BHF approach are identical to
those of HF. Indeed, this correction increases the calculated binding
energy to -4.01 MeV. This number is in reasonable agreement with
self-consistent BHF calculations performed for $^{16}$O using the same
interaction \cite{carlo}. However, as the single-particle states of
BHF are more bound than the single-particle states obtained in HF, the
gain in the binding energy from HF to BHF is accompanied by a reduction
of the calculated radius of the nucleon distribution. This is the well-known
phenomenon of the so-called ``Coester band'' in finite nuclei \cite{carlo},
which plagues microscopic attempts to calculate ground-state properties of
nuclear systems already for a very long time \cite{coester}.

The inclusion of the 2h1p contributions to the self-energy in the complete
calculation reduces the absolute values of the quasihole energies (compare
BHF and ``Total'' in Tab.\ref{tab:kol}). This is to be expected from our
discussion following Eq. (\ref{eq:disper2}). Despite this reduction of the
quasihole energies, however, the total binding energy is increased as
compared to BHF. This increase of the binding energy is mainly due to the
continuum part of the spectral function. Comparing the various parts of the
``Koltun sumrule'' of Eq. (\ref{eq:koltun}) one finds that only 37 percent
of the total energy is due to the quasihole part of Eq. (\ref{eq:koltun}).
The dominating part (63 percent)
results from the continuum part of the spectral functions
although this continuum part only ``represents 11 percent of the nucleons''
(see Tab.\ref{tab:noc}).

The calculation of the radius, however, is dominated by the quasihole
contribution to the density. As the quasihole terms have reduced energies
as compared to BHF, it is plausible that the calculated radius increases
in the total
calculation as compared to BHF. Therefore the inclusion of 2h1p terms
increases the calculated binding energy and radius, moving the results
for the ground state off the Coester band into the direction of the
experimental data. This effect is large enough to explain the discrepancy
obtained between the experimental data and the results of
microscopic Dirac-Brueckner-Hartree-Fock calculations including relativistic
effects \cite{fritz}. We note that inclusion of three-body forces in
variational calculations for $^{16}$O also yields very good results for the
binding energy \cite{pwp}.

\section{Conclusions}

An attempt has been made to derive the spectral function and the momentum
distribution from a realistic OBE interaction directly for a finite nucleus
without the assumption of a local density approximation. The correlations
taken into account beyond the Hartree-Fock approximation yield a strong
enhancement of the momentum distribution at high momenta. It is demonstrated
that this enhancement originates from the spectral function at large
negative energies and therefore should be observed in nucleon knockout
reactions with large energy transfer leaving the final nucleus at an
excitation energy of about 100 MeV.

The enhancement of the high-momentum components is weaker as obtained in
studies of nuclear matter. This difference may be due to the different
interactions employed (unfortunately no nuclear matter result is available
for the OBE interaction used here) or due to approximations used in the
calculation for the finite system. Therefore further studies of these
approximations (poor treatment of low-energy excitations, the self-energy
of the nucleons is calculated in a perturbative way including terms up to
second order in $G$) is required before conclusions about the validity of
the Local Density Approximation relating the results of nuclear matter
to those of finite nuclei can be drawn.

The resulting Green function is also used to determine the total energy
and the radius of the nucleon distribution. It is demonstrated that the
inclusion of 2-hole 1-particle contributions to the self-energy of the
nucleon yields an enhancement of the calculated binding energy per nucleon
($\approx $ 1 MeV) and an increase of the radius ($\approx$ 0.05 fm) for
$^{16}$O as compared to the Brueckner-Hartree-Fock approach. This could be
sufficient to explain the discrepancy remaining between experimental data
and microscopic Dirac-BHF calculations \cite{fritz}.

This research project has partially been supported by SFB 382 of the
"Deutsche Forschungsgemeinschaft", DGICYT, PB92/0761
(Spain), the EC-contract CHRX-CT93-0323, and the U.S. NSF under Grant No.
PHY-9307484. One of us (H.M.) is pleased
to acknowledge the warm hospitality at the Facultad de Fisica, Universitad de
Barcelona, and the support by the program for Visiting Professors of this
university.

\begin{table}[h]
\caption{Distribution of nucleons in $^{16}$O.
Listed are the total occupation number $\hat n$ for various
partial waves (see Eq.(34))
but also the contributions from the quasihole ($\hat n^{qh}$) and the
continuum part ($\hat n^c$) of the spectral function, separately.
The continuum part is split further into contributions originating
from energies $E$ below -150 MeV ($n^c(E<-150)$) and from energies
below -100 MeV.
The last line shows the sum of particle numbers for all partial waves
listed.}
\label{tab:noc}
\begin{center}
\begin{tabular}{c|rrrr|r}
&&&&&\\
$lj$&\multicolumn{1}{c}{$\hat n^{qh}$}&\multicolumn{1}{c}{$\hat n^c(E<-150)
$}&\multicolumn{1}{c}{$\hat n^c(E<-100)$}&
\multicolumn{1}{c|}{$\hat n^c$}&\multicolumn{1}{c}{$\hat n$}
 \\
&&&&&\\ \hline
&&&&&\\
$s_{1/2}$ & 3.120 & 0.033 & 0.244 & 0.624 & 3.744 \\
$p_{3/2}$ & 7.314 & 0.032 & 0.133 & 0.332 & 7.646 \\
$p_{1/2}$ & 3.592 & 0.026 & 0.086 & 0.173 & 3.764 \\
&&&&&\\
$d_{5/2}$ & 0.0 & 0.033 & 0.106 & 0.234 & 0.234 \\
$d_{3/2}$ & 0.0 & 0.036 & 0.108 & 0.196 & 0.196 \\
$f_{7/2}$ & 0.0 & 0.025 & 0.063 & 0.117 & 0.117 \\
$f_{5/2}$ & 0.0 & 0.032 & 0.084 & 0.140 & 0.140 \\
&&&&&\\
$\sum$ & 14.025 & 0.217 & 0.824 & 1.816 & 15.841 \\
&&&&&\\
\end{tabular}
\end{center}
\end{table}
\begin{table}[h]
\caption{Groundstate properties of $^{16}$O. Listed are the energies
$\epsilon$ and kinetic energies $t$ of the quasihole states (qh) and the
corresponding mean values for the continuum contribution (c), normalized to 1,
for the various partial waves. Multiplying the sum: $1/2(t+\epsilon )$ of these
mean values with the corresponding particle numbers of Tab.I, one obtains
the contribution $\Delta E$ to the energy of the ground state (see the
Koltun sumrule Eq. (31)). Summing up all these contributions and dividing
by the nucleon number yields the
energy per nucleon $E/A$. Furthermore we give the radius for nucleon
distribution $\left\langle r \right\rangle$,
calculated from the square root of Eq. (29). Results are presented for
the Hartree-Fock (HF), Brueckner-Hartree-Fock (BHF) and the complete
calculation (Total). The particle numbers for the qh states in HF and BHF
are equal to the degeneracy of the states, all other occupation numbers
are zero. The results for the radii are given in fm, all other entries in MeV}
\label{tab:kol}
\begin{center}
\begin{tabular}{c|rrr|rrr|rrr}
&&&&&&&&&\\
&\multicolumn{3}{c|}{HF} &\multicolumn{3}{c|}{BHF} &\multicolumn{3}{c}
{Total} \\
$lj$&\multicolumn{1}{c}{$\epsilon$}&\multicolumn{1}{c}{$t$}&
\multicolumn{1}{c|}{$\Delta E$}
&\multicolumn{1}{c}{$\epsilon$}&\multicolumn{1}{c}{$t$}&
\multicolumn{1}{c|}{$\Delta E$}
&\multicolumn{1}{c}{$\epsilon$}&\multicolumn{1}{c}{$t$}&
\multicolumn{1}{c}{$\Delta E$}\\
&&&&&&&&&\\
\hline
&&&&&&&&&\\
$s_{1/2}$ qh & -36.91 & 11.77 & -50.28 & -42.56 & 11.91 & -61.30 & -34.30
& 11.23 & -35.98 \\
$s_{1/2}$ c & &&&&&& -90.36 & 17.09 & -22.89 \\
$p_{3/2}$ qh & -15.35 & 17.62 & 9.08 & -20.34 & 18.95 & -5.59 & -17.90 &
18.06 & 0.37 \\
$p_{3/2}$ c  & & & & & & & -95.19 &
35.19 & -9.96\\
$p_{1/2}$ qh & -11.46 & 16.63 & 10.34& -17.07 & 18.46 &  2.76 & -14.14 &
17.19 & 5.47 \\
$p_{1/2}$ c  & & & & & & & -103.62&
35.94 & -5.84\\
$l>1$ c  & & & & & & & -98.87 &
63.17 & -12.27 \\
&&&&&&&&&\\
\hline
&&&&&&&&&\\
$E/A$&\multicolumn{3}{c|}{-1.93} &\multicolumn{3}{c|}{-4.01}
&\multicolumn{3}{c}
{-5.12} \\
$\left\langle r \right\rangle$&\multicolumn{3}{c|}{2.59}
&\multicolumn{3}{c|}{2.49} &\multicolumn{3}{c}
{2.55} \\
&&&&&&&&&\\
\end{tabular}
\end{center}
\end{table}
\clearpage
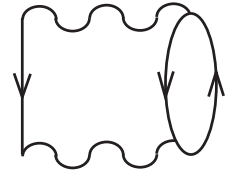
\begin{figure}
\caption{Graphical representation of the Hartree-Fock (a), the 2-particle
1-hole (2p1h, b) and the 2-hole 1-particle contribution (2h1p, c) to the
self-energy of the nucleon}
\label{fig:diag}
\end{figure}
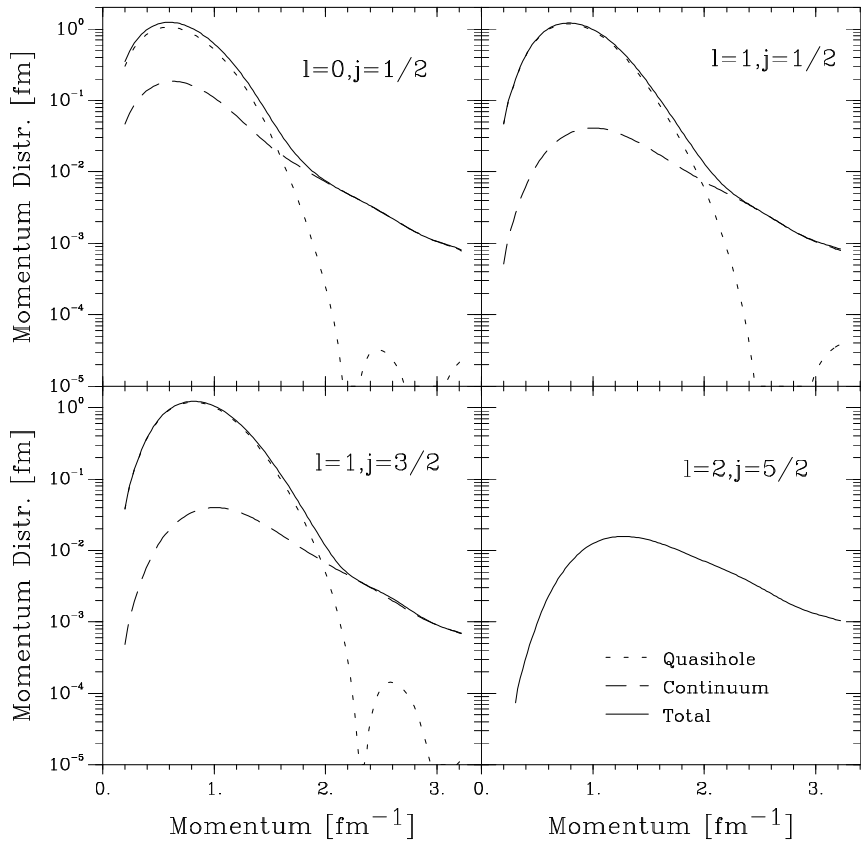
\begin{figure}
\caption{Momentum distribution for different partial waves in $^{16}$O
(see Eq. (2)) obtained by integrating the spectral function $S_{lj}(k,E) $
(see Eq. (3)).
The momentum distribution
is the sum of the quasihole contribution (dashed curve)
and the continuum contribution (dotted curve). All functions are normalized
such that
$\int dk\ n(k)\ =\ 1$ if $S(k) $ refers to an orbit which is mostly
occupied.}
\label{fig:specpw}
\end{figure}
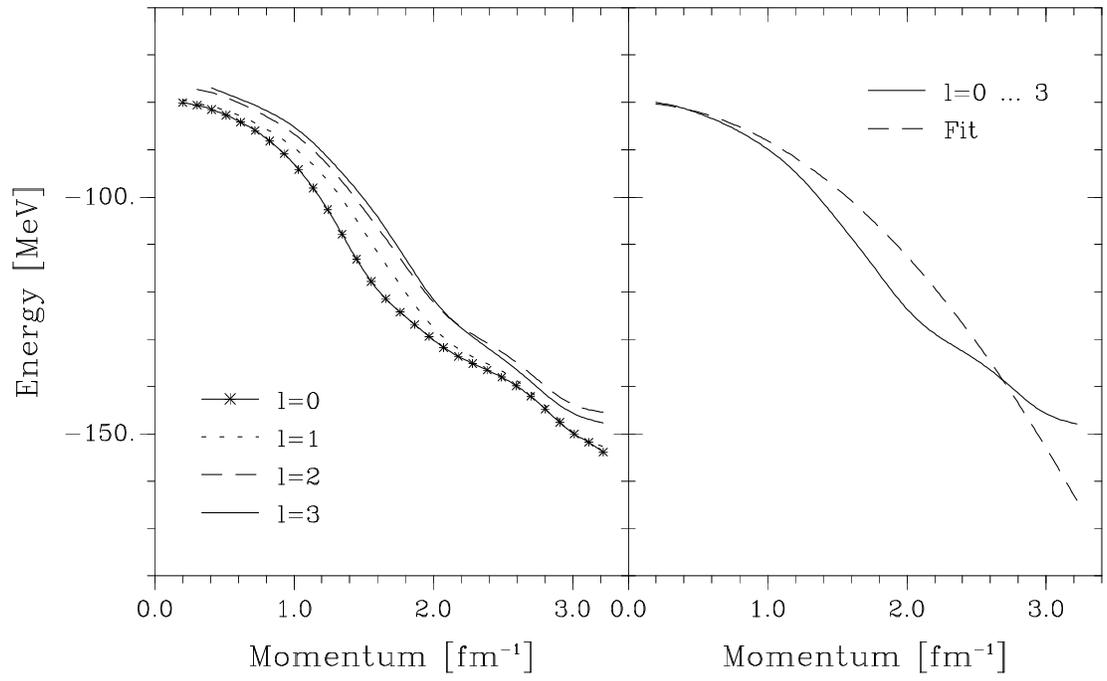
\begin{figure}
\caption{Mean value for the energy of the 2h1p continuum as a function
of the momentum $k$. The left part of this figure shows results for
${\cal E}_lj$ (see Eq. (32)) in various partial waves. In the right part
of the figure
the mean value averaged for the various $l$ and $j$ (see Eq. (33)) is
displayed. For comparison this part also includes a simple parametrization
in terms of $-k^2 /(2m^*) - C$ with $m^*$ = 2400 MeV and $C$ = 80 MeV.}
\label{fig:emean}
\end{figure}
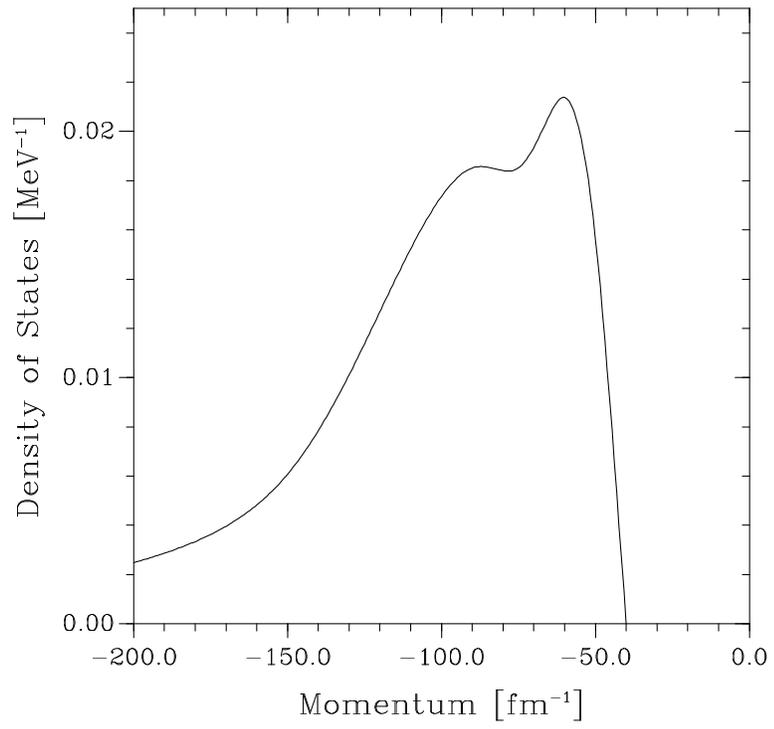
\begin{figure}
\caption{Density of states or total occupation probability of the 2h1p
continuum as a function
of the energy $E$ (see Eq. (35)). The normalization of this distribution
is such that the integration over the energy yields the total particle
number of 1.816 (see Tab.I) in the continuum.}
\label{fig:nce}
\end{figure}
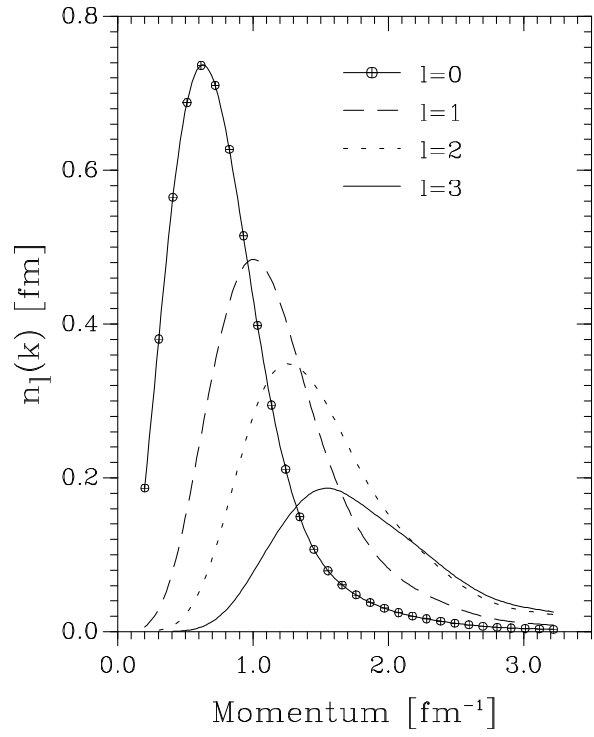
\begin{figure}
\caption{The momentum distribution for various orbital angular momenta.
These distribution account for the different $j$, include
the degeneracy factors $2(2j+1)$, and are normalized
in such a way that $\int dk n_l(k)$ yields the total number of particles
with orbital angular momentum $l$ in the 2p1h continuum.}
\label{fig:npw}
\end{figure}
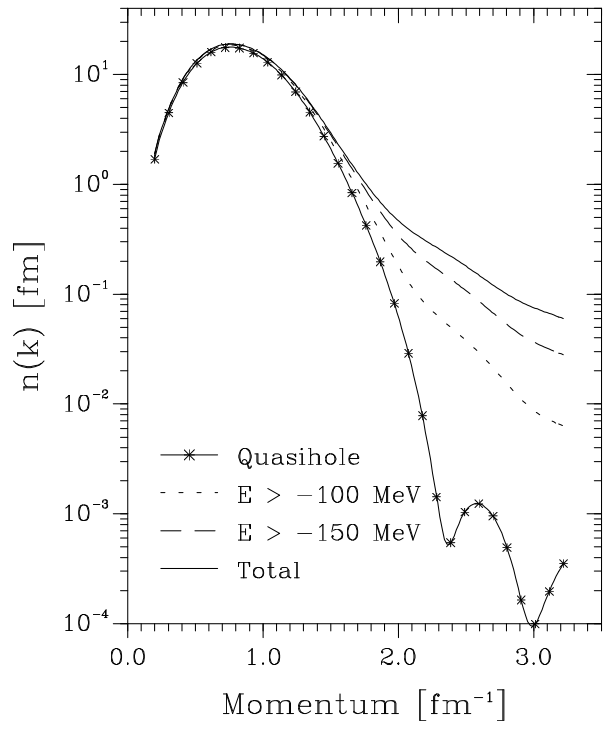
\begin{figure}
\caption{The total momentum distribution of $^{16}$O. Shown are also the
quasihole contribution and the results obtained with various energy
cut-offs in the integration of the spectral functions}
\label{fig:nto1}
\end{figure}
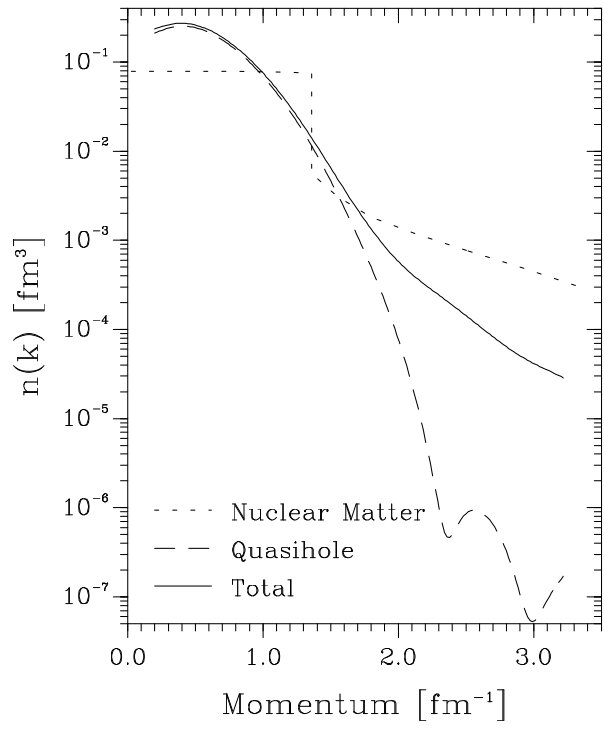
\begin{figure}
\caption{The total momentum distribution obtained in the present investigation
for $^{16}$O employing the OBE potential $B$ of Ref.\ 24 is compared to the
momentum distribution obtained for the Reid soft-core potential in nuclear
matter [14]. In this figure the momentum distributions are normalized
in such a way that $\int d^3k\, n(k)$ yields 1.}
\label{fig:nto2}
\end{figure}
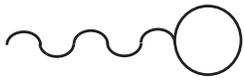
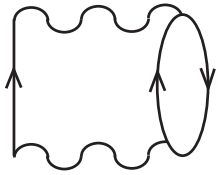
\end{document}